\documentclass[referee]{raa}            
\usepackage{graphicx,times}             
\usepackage{amssymb,amsmath,multicol,verbatim,eurosym,epsfig,latexsym}
\usepackage[subfigure]{tocloft}
\usepackage{natbib}
\bibpunct{(}{)}{;}{a}{}{,}
\usepackage{float}
\usepackage[pagebackref=true]{hyperref}
\usepackage{geometry}
\newcommand{\orcid}[1]{\href{https://orcid.org/#1}{\includegraphics[width=10pt]{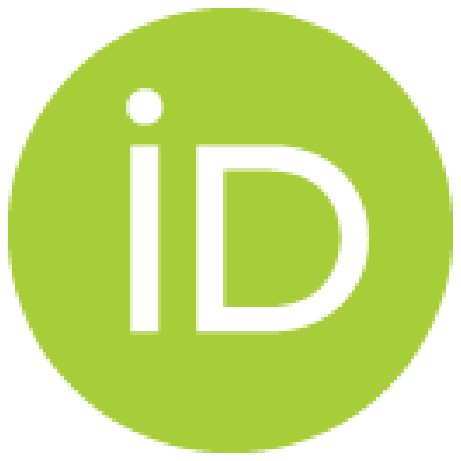}}}

\usepackage{kvsetkeys}
\usepackage[dvipsnames]{xcolor}

\renewcommand{\d}{\mathrm{d}}

\newcommand{\bea}{\begin{eqnarray}}
\newcommand{\eea}{\end{eqnarray}}
\newcommand{\rund}[1]{\left(#1\right)}

\begin{document}

   \title{The chromatic Point Spread Function of weak lensing measurement in Chinese Space Station survey Telescope
}
   \volnopage{Vol.0 (20xx) No.0, 000--000}      
   \setcounter{page}{1}          

   \author{
     Quanyu Liu\inst{1}\orcid{0000-0003-4875-2992}\thanks{Email: muttonshashlik@outlook.com}, Xinzhong Er\inst{1}\thanks{Email: phioen@163.com},
     Chengliang Wei\inst{2},
     Dezi Liu\inst{1},
     Guoliang Li\inst{2},
     Zuhui Fan\inst{1},
     Xiaobo Li\inst{3},
     Zhang Ban\inst{3},
     Dan Yue\inst{4}
   }
   \institute{
South-Western Institute for Astronomy Research, Yunnan University, Kunming 65000, P.R.\,China\\
\and
Purple Mountain Observatory, Chinese Academy of Sciences, Nanjing, Jiangsu, 210023, China\\
\and
Changchun Institute of Optics, Fine Mechanics and Physics, Chinese Academy of Sciences, Changchun, 130033, China \\
\and
College of Physics, Changchun University of Science and Technology, Changchun 130022, China \\
\vs\no
{\small Received~~20xx month day; accepted~~20xx~~month day}}

\abstract{The weak gravitational lensing is a powerful tool in modern cosmology. To accurately measure the weak lensing signal, one has to control the systematic bias to a small level. One of the most difficult problems is how to correct the smearing effect of the Point Spread Function (PSF) on the shape of the galaxies. The chromaticity of PSF for a broad-band observation can lead to new subtle effects. Since the PSF is wavelength dependent and the spectrum energy distributions between stars and galaxies are different, the effective PSF measured from the star images will be different from that smears the galaxies. Such a bias is called colour bias. We estimate it in the optical bands of the Chinese Space Station Survey Telescope from simulated PSFs, and show the dependence on the colour and redshift of the galaxies. Moreover, due to the spatial variation of spectra over the galaxy image, there exists another higher-order bias, colour gradient bias. Our results show that both colour bias and colour gradient bias are generally below $0.1$ percent in CSST. Only for small-size galaxies, one needs to be careful about the colour gradient bias in the weak lensing analysis using CSST data.
}
   \authorrunning{Liu et al. }  
   \titlerunning{Chromatic PSF in CSST}  

   \maketitle
%
\section{Introduction} 
\label{sect:intro}

The light from distant objects is deflected by the gravitational potential of the massive objects along their path to us, which is referred to as gravitational lensing \citep[e.g.][]{1991MNRAS.251..600B,2001PhR...340..291B}.In the limit of very weak deflection, i.e. without striking phenomena such as multiple images or arcs, lensing is referred to as ''weak lensing'' \citep[e.g.][]{2015RPPh...78h6901K}. The images of distant galaxies are weakly distorted by the tidal effect of the gravitational potential, i.e. lensing shear. The resulting correlations in the shapes can be related directly to the statistical properties of the mass distribution in the universe, thus the weak lensing by large-scale structure, or cosmic shear, has been identified as a powerful tool for cosmology, and has been demonstrated by several observations, e.g. the COSMOS survey \citep{2010A&A...516A..63S}, the CFHTLenS survey \citep{2012MNRAS.427..146H} etc. More recent studies yield competitive constraints on cosmological parameters \citep[e.g.][]{2018PhRvD..98d3526A,2021A&A...645A.104A,2020PASJ...72...16H}. Thus several ongoing or future missions are designed with weak lensing as a primary science driver, such as the Vera Rubin Observatory Legacy
Survey of Space and Time \citep[the LSST][]{2009arXiv0912.0201L,LSST2019}, the ESA space-borne telescope $Euclid$ \citep{2011arXiv1110.3193L}, the Roman Space Telescope \citep[WFIRST][]{Roman2015} and the Chinese Space Station Survey Telescope \citep[CSST][]{Zhan2011,Zhan2018}.

The measurement of the cosmic shear requires stringent control of the systematic effects since the statistical error of large weak lensing surveys becomes less important \citep[e.g.][]{2018ARA&A..56..393M}. In the past few decades, significant progress has been achieved on the shear measurements \citep[e.g.][]{1995ApJ...449..460K,1997ApJ...475...20L,1999A&A...352..355K,2002AJ....123..583B,2011JCAP...11..041Z,2013MNRAS.429.2858M,2021A&A...656A.135H}. One of the most dominant sources of measurement bias is the smearing of the images by the Point Spread Function (PSF). Precise modelling of the PSF is crucial and difficult. In reality, people usually estimate the PSF by measuring the shapes of star images \citep[e.g.][]{2004MNRAS.347.1337H,2021MNRAS.501.1282J}, and construct the PSF model over the whole field of view. The implicit assumption is that the PSF affecting stars and galaxies is locally the same. However, it has been noticed that the PSF has a dependence on the observing wavelength. The measured star images over a wide band can provide an ``effective'' PSF, which is weighted by the Spectral Energy Distribution (SED) of the star. Once the SEDs of the galaxies are different from that of stars, the assumption that the effective PSF from a star is the same as that of a galaxy will be violated. Such a deviation is called colour bias. \citet{2010MNRAS.405..494C} firstly proposed such an issue and discussed the impact on the shear measurements in a diffraction-limited telescope. They find such kind of a bias can be calibrated in cases of (i) the stars have the same colour as the galaxies; (ii) estimation of the galaxy SED using multiple colours and a PSF model of PSF using the optical design of the telescope. \citet{2015ApJ...807..182M} study the impact of the colour bias on shape measurements of two atmospheric chromatic effects for ground-based surveys, the Dark Energy Survey (DES) and the LSST. \cite{2018MNRAS.477.3433E} explores various approaches to determine the effective PSF using broad-band data. They also study the correlations between photometric redshift and PSF estimates that arise from the use of the same photometry, considering the Euclid mission as a reference. \cite{2018MNRAS.479.1491C} measures the wavelength dependence of the PSF size in the Hyper Suprime-Cam (HSC) Subaru Strategic Program (SSP) survey, and constructs a PSF size model as a function of the wavelength. They proposed a power-law model and tried to calibrate the colour bias to fulfil the error budget proposed by \cite{2015ApJ...807..182M}. \cite{2012PASP..124.1113P} discuss how differential chromatic refraction can bias shear measurements in LSST by introducing a SED-dependent elongation of the PSF along the elevation vector.

Besides the colour bias, the galaxy shapes can differ significantly across filters. In other words, the SED of the galaxies varies spatially. Since in the weak lensing, the signature is achromatic, it has been proposed to measure cosmic shear from multiple filters \citep{2008JCAP...01..003J}. However, a higher-order systematic bias in shear measurement can arise within a filter due to such an effect, which is called colour gradient bias \citep[CG bias for short][]{2012MNRAS.421.1385V,2013MNRAS.432.2385S,2020ApJ...888...23K}. They estimate the CG bias with the chromatic PSF for the wide band images of Euclid and LSST, showing that the bias has to be taken into account for precise shear measurement. \cite{2018MNRAS.476.5645E} performs analysis of CG bias using real data taken by the Hubble Space Telescope, and demonstrates that the CG bias can be calibrated by images from two narrower bands and presents its dependence on the galaxy properties, such as colour, galaxy size etc.

The stage-IV cosmological surveys target high-precision weak lensing measurements for over a billion galaxies. The shrinking statistical error causes all the systematic biases prominent. Thus, even the small, higher-order systematic bias, such as colour bias and CG bias need to be carefully estimated and controlled. In this work, we study the two biases due to the chromatic PSF effect in CSST weak lensing images. The basic formulae are given in Sect.\,\ref{sect:formula}. The estimates of colour bias and CG bias are given in the following Sects.\,\ref{sect:colourbias} and \ref{sect:CGbias}. We discuss our results in the end.

\section{The basic formulae}
\label{sect:formula}
We follow the conventional notations of gravitational lensing \citep[e.g.][]{2001PhR...340..291B}.
We introduce the angular coordinates ${\theta} =(\theta_1,\theta_2)$ on the lens plane, which is perpendicular to the line of sight. For an image of a galaxy or a star, we denote the photon brightness distribution of the image at each position $\theta$ and wavelength $\lambda$ by $I(\theta,\lambda)$. The resulting image  observed with a PSF $P(\theta,\lambda)$ in a bandwidth $\Delta \lambda$ is given by
\be
I^{\rm obs}(\theta)=\int_{\Delta \lambda} \d \lambda\, I(\theta,\lambda) * P(\theta,\lambda),
\ee
where $*$ denotes convolution, and
$I(\theta;\lambda)$ is the brightness of the pre-convolution image, $I(\theta,\lambda)=\lambda S(\theta,\lambda) T(\lambda)$. $S(\theta,\lambda)$ is the SED of source at position $\theta$, and $T(\lambda)$ is the transmission function of the filter. The size of a star image usually is smaller than the pixel scale and can be considered as a delta function before the PSF smearing. Thus the observed star image can be used to estimate the PSF. The observed star image can be written as an integration of PSF at each wavelength and weighted by the star SED $S^{\rm star}(\lambda)$,
\be
P^{\rm star}(\theta) = {\int \d\lambda P(\theta,\lambda) T(\lambda) S^{\rm star}(\lambda) \lambda},
\label{eq:define_starPSF}
\ee
which is also called the {\it effective PSF} in the shear measurements. Henceforth the PSF means effective PSF if we do not mention it. The analogous PSF which smears the galaxy and is weighted by the SED of the galaxy is named ''galaxy PSF''
\be
P^{\rm gal}(\theta) = {\int \d\lambda P(\theta,\lambda) T(\lambda) S^{\rm gal}(\lambda) \lambda}.
\label{eq:define_galPSF}
\ee

Apparently, $P^{\rm gal}$ cannot be measured directly.
If the SED of the star has the same shape as that of the galaxy SED, then the PSF estimated from the star image can be used for the shear measurement. Otherwise, the difference between the effective PSF using star SED and that using galaxy SED will introduce a bias in the shear measurements, which is called '{\it colour bias}'. We make following simplifications in our study of colour bias: (1) we ignore the difference of the PSF over the FOV, i.e. at the star position and that at the galaxy position. (2) We define the colour bias by the difference between the two effective PSFs, i.e. $P^{\rm star}$ and $P^{\rm gal}$. The measurement error in the cosmic shear is discussed later for only one particular case. (3) galaxy has a spatially uniform SED, i.e. the morphology of the galaxy is not taken into account in the analysis of colour bias. In the study of colour gradient bias (Sect.\ref{sect:CGbias}), the spatial variation of SED or the morphology of the galaxy is considered.

Several methods have been proposed and adopted in the shear measurements. First one is based on the brightness moments of the galaxy images,\citep[e.g.][]{1995ApJ...449..460K,1997ApJ...475...20L,1998ApJ...504..636H}.
Second one makes use of the model fitting of the galaxy shape \citep[e.g.][]{1999A&A...352..355K,2002AJ....123..583B,2002sgdh.conf...29R,2010MNRAS.404..458V}. Some disadvantages have been found in these methods\citep[e.g.][]{2011MNRAS.414.1047Z,2017ApJ...841...24S,2018ARA&A..56..393M}, and more new measurement methods have been proposed \citep[e.g.][]{2015JCAP...01..024Z,2019AAS...23334909E,2020MNRAS.491.5301S,2021arXiv210409970T}. Each method will suffer different colour and colour gradient biases. To simplify the calculation, we adopt the brightness moment method to quantify the shape of PSF and galaxy throughout this paper. Other kinds of measurement noise or bias are not considered either. Therefore, we define the difference of second order brightness moment between $P^{\rm star}$ and $P^{\rm gal}$ as the colour bias. The second order brightness moment is written as
\be
Q_{ij} \equiv\frac{\int d^2 \theta \,\theta_i\,\theta_j\,W(\theta) \int d\lambda\,I(\theta;\lambda) * P(\theta;\lambda)}{\int d^2\theta\, W(\theta)\int d\lambda\,I(\theta;\lambda) * P(\theta;\lambda)},
\ee
where $W(\theta)$ is the weight function to reduce the noise in real measurement, $i,j$ indicates two directions on the sky. $I(\theta;\lambda)$ is the delta function when we calculate the moments of the stellar image. Then the size and the ellipticity of the image can be calculated from
\begin{align}
&R=\sqrt{Q_{ii} + Q_{jj}},\quad \epsilon_1 = \dfrac{Q_{ii}-Q_{jj}}{Q_{ii}+Q_{jj}},\quad 
\epsilon_2 = \dfrac{2Q_{ij}}{Q_{ii}+Q_{jj}},
\end{align}
where $R$ represent the size, $\epsilon_1$ and  $\epsilon_2$ are two components of ellipticity $\epsilon=(\epsilon_1^2 + \epsilon_2^2)^{1/2}$. We estimate the colour bias for the size $R$ and the module of the ellipticity $\epsilon$ separately, which is defined as
\be
b_c\equiv\frac{R^{\rm gal}-R^{\rm star}}{R^{\rm star}},\; {\rm for\; size\; of\; PSF;} 
\qquad 
b_c\equiv\frac{\epsilon^{\rm gal}-\epsilon^{\rm star}}{\epsilon^{\rm star}}\; {\rm for\; ellipticity\; of\; PSF,}
\label{eq:colourbias}
\ee
where the superscript 'gal' or 'star' indicates that the quantities are calculated from the effective PSF  using the SED of galaxy or star respectively.

Moreover, the shape of the galaxy varies over the wavelength (or the spectral energy distribution of galaxy varies spatially), i.e. the colour gradient. The shape measurement without taking into account such a kind of effect can induce another bias as well, which is called '{\it colour gradient bias}' (CG bias for short).
The multiplicative bias induced by the colour gradient in shear measurement is defined by 
\be
b_{CG} = \frac{\hat\gamma_i-\gamma_i}{\gamma_i},\quad i=1,2,
\ee
where the subscript '$i$' indicates one of two components of the shear. $\gamma_i$ is the true shear and $\hat{\gamma_i}$ is the estimate of shear without taking into account of colour gradient. One can find more details on the CG bias in\citet{2013MNRAS.432.2385S,2018MNRAS.476.5645E}.

\section{Colour bias and calibration}
\label{sect:colourbias}

\subsection{Simulation of colour bias}
\begin{figure}
\centerline{	\includegraphics[width=1\textwidth]{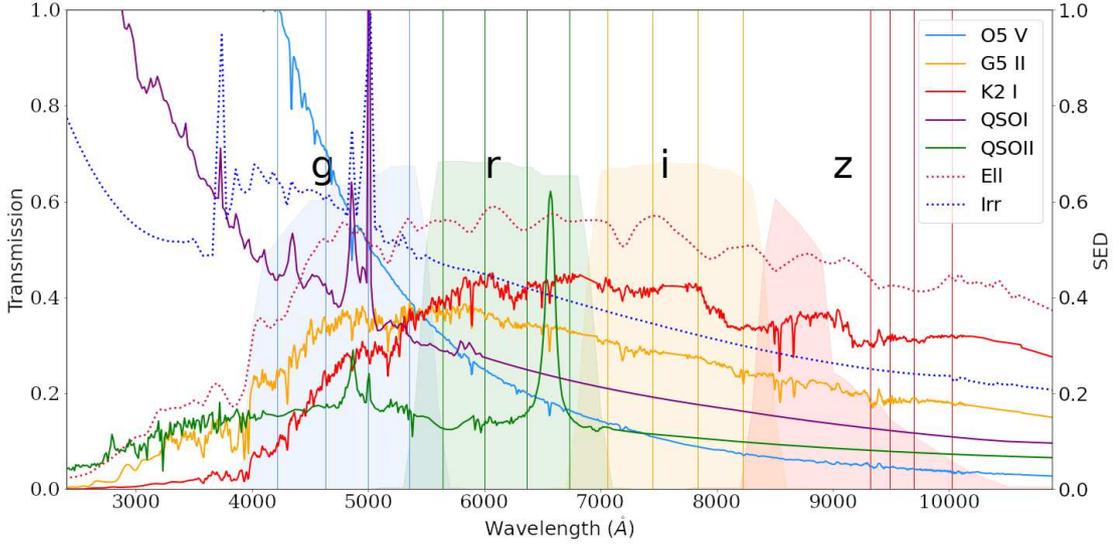}}
\caption{ The transmission function of four filters from CSST (shaded region), and the spectra energy distribution (SED) of objects that are used in our simulation. The crimson dotted line and blue dotted line present the SED of an elliptical galaxy (Ell) and an irregular galaxy (Irr). The dodger blue, orange, and red solid lines present three different types of star SEDs, which are O5 V, G5 II, and K2 I respectively. The purple and green solid lines indicate two different types of QSO SED (QSOI and QSOII). The vertical lines indicate the wavelengths where we have the simulated PSFs.}
\label{fig:filters&seds}
\end{figure}

\begin{figure}
\centering	
\includegraphics[width=0.9\textwidth]{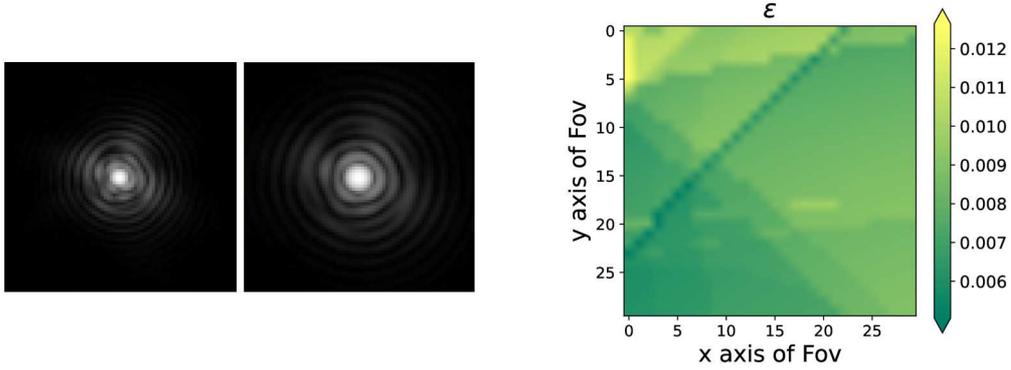}
\caption{Left panel: two examples of simulated PSF of CSST at $4630\mathring{A}$ in the $g$ band (left) and $7500\mathring{A}$ in the $i$ band (right). The intensity is given in the log scale in arbitrary units. Right panel: the spatial distribution of the PSF ellipticity on the FOV at wavelength $7830\mathring{A}$ in the $i$ band.}
\label{fig:showPSF}
\end{figure}

\begin{figure*}
\centerline{
\includegraphics[width=0.5\textwidth]{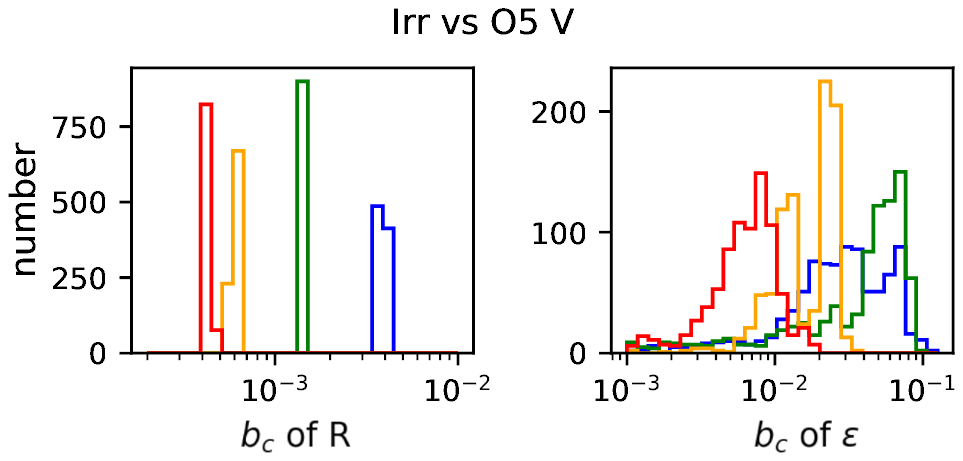}
\includegraphics[width=0.5\textwidth]{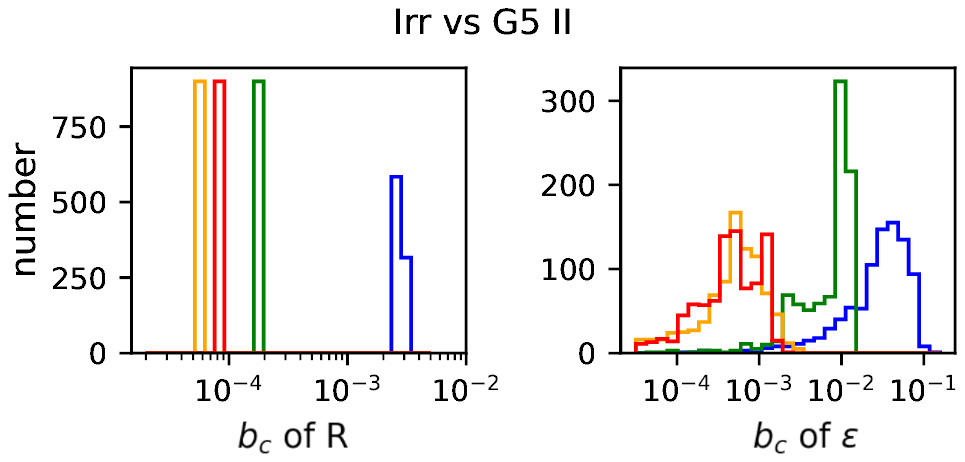}}
\centerline{
\includegraphics[width=0.58\textwidth]{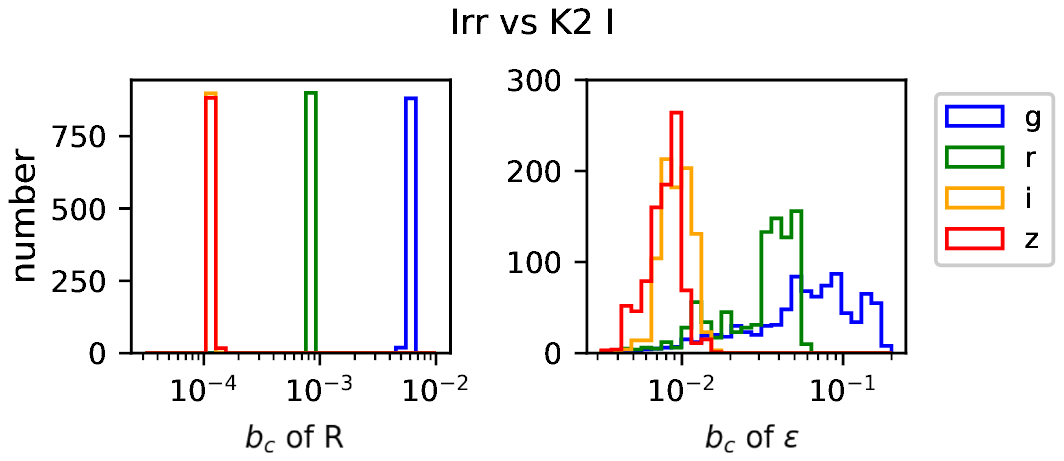}}
\centerline{
\includegraphics[width=0.5\textwidth]{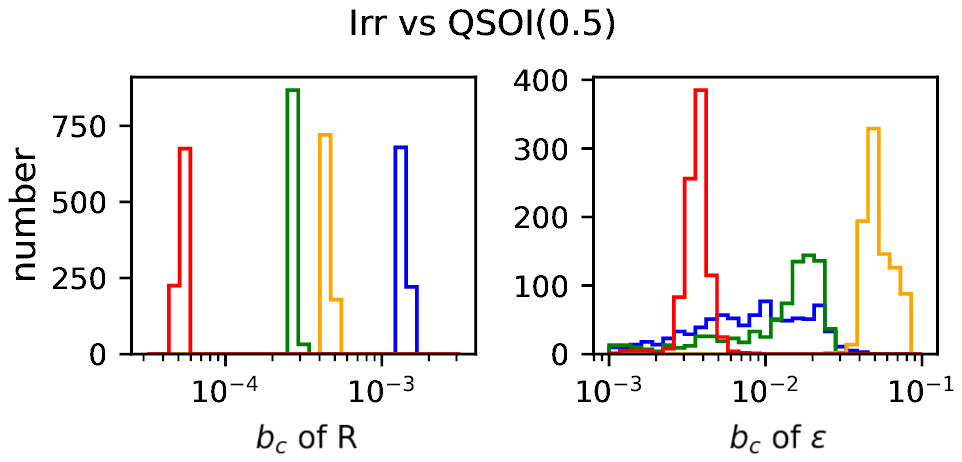}
\includegraphics[width=0.5\textwidth]{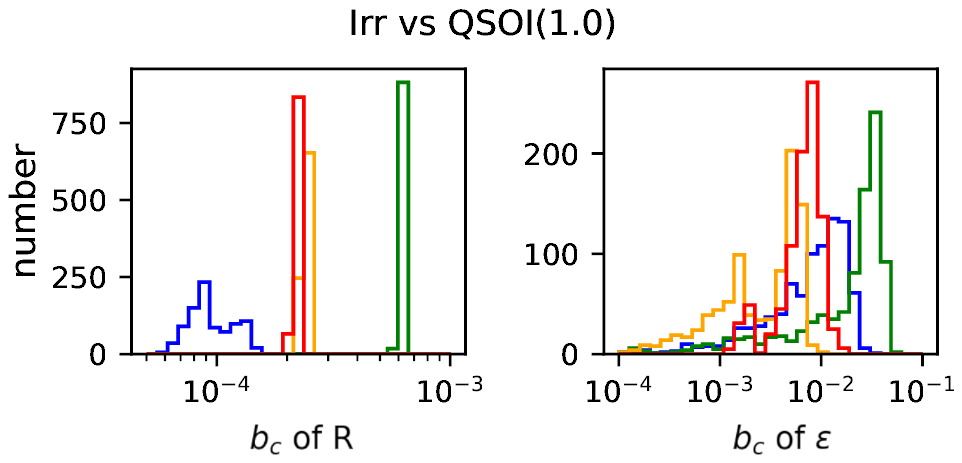}}
\centerline{
\includegraphics[width=0.5\textwidth]{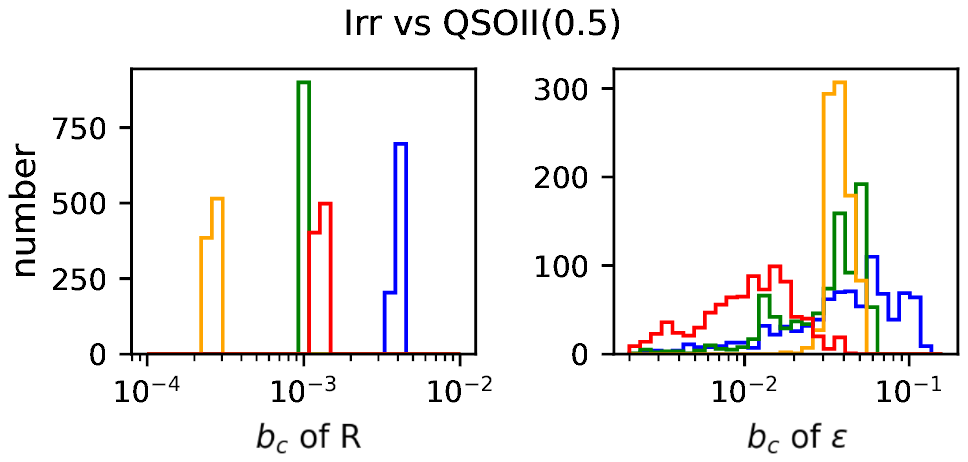}
\includegraphics[width=0.5\textwidth]{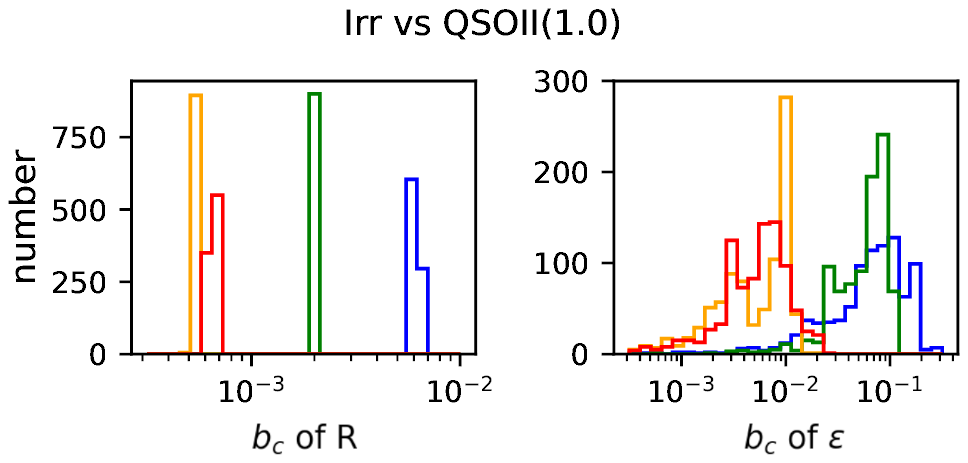}}    
\caption{The histogram of the colour bias in the CSST $g, r, i, z$ band using blue, green, orange and red colours respectively. The results between Irr and three types of stars (O5 V, G5 II, K2 I), as well as two types of QSOs (QSOI and QSOII), are shown. We use typical redshifts for QSOI and QSOII which is 0.5 (left) and 1.0 (right). In each panel, from left to right is the colour bias of $R$ and $\epsilon$ respectively.}
\label{fig:show colour bias}
\end{figure*}

In order to evaluate the colour bias of CSST, we compare $R$ and $\epsilon$ of the simulated PSF weighted by SED from stars, galaxies or Quasars. Four bands ($g,r,i,z$) from CSST are studied, with the predicted transmissions of each band given by the shaded regions in Fig.\,\ref{fig:filters&seds}.
The simulated PSFs are obtained from the design of optics, by the CSST image simulation team. To generate a set of realistic PSFs and to account for the impact of the optical system on image quality, an optical emulator has been developed to simulate high-fidelity PSFs of CSST. The optical emulator of CSST is based on six different modules to simulate the optical aberration due to mirror surface roughness, fabrication errors, CCD assembly errors, gravitational distortions, thermal distortions, installation, and adjustment errors. Moreover, two dynamical errors, due to micro-vibrations and image stabilisation, are also included in the simulated PSF. We use one of the modules which includes all the effects and is close to the realistic case.

The PSFs are simulated on a $30\times30$ grid uniformly on the whole Field-Of-View (FOV). Each stamp of PSF simulation is comprised of $512\times512$ pixels with pixel scale $0.0185$ arcsec. At each grid, there are simulated PSF images at four different wavelengths in each band. We do not find a strong variation of the size of the PSF over the FOV, while the ellipticity of the PSF varies rapidly (right panel of Fig.\,\ref{fig:showPSF}). There are variations of the PSF at different wavelengths. The mean value of PSF ellipticity and size for the longest and shortest wavelength in our simulations is $4.71\times10^{-3}$, $6.43\times10^{-3}$ and $24.03$, $26.24$ pixels respectively. Here we use the modulus of ellipticity. The wavelengths are indicated by the vertical lines in Fig.\,\ref{fig:filters&seds}. We show an example of the simulated PSF in the left panels of Fig.\,\ref{fig:showPSF}.

We choose two kinds of SEDs for the reference galaxies, three for the stars and two for the QSOs. The galaxy SED templates from \cite{1980ApJS...43..393C}: an elliptical galaxy which has a relatively red colour and an irregular galaxy which has a blue colour. The two distinguishing spectra can give us an up limit of the colour and colour gradient bias. For short, we dub the two galaxies as Ell and Irr respectively throughout the paper. The SED templates of stars are taken from \cite{1998PASP..110..863P}: the stars of type O5 V, G5 II and K2 I. O5 V have a blue colour, K2 I has a red colour and G5 II is somewhere in between. In addition, QSOs have point-like image morphology and can be misclassified as stars for the PSF estimation. We study the colour bias from QSO SED as well. The SED templates of QSOs are adopted from SWIRE library \cite{2007ApJ...663...81P}: a Type-1 QSO (QSOI) and a Type 2 QSO (QSOII). 
 The probability of mis-classification of QSO to star is difficult to estimate and depends on the observations. For example, one can reach high precision with spectra. In \cite{2022arXiv221205868V}, the classification of galaxy-star-QSO using the J-PLUS DR3 has been carried out. One million stars and 0.2 million QSOs have been identified. The result shows that using twelve photometric bands, they can reach a precision of 0.94 for QSOs. Given the high number density of QSOs, and even higher at high redshift, it is necessary to include the contamination of the QSOs. All the SEDs that we used are shown by the curves in Fig.\,\ref{fig:filters&seds}, in which all the SEDs are re-scaled arbitrarily for better visibility. 

Because we have only four simulated PSFs on different wavelengths in each band, there exist further uncertainties in our estimates of the colour bias. For example, as one can see from Fig.\,\ref{fig:filters&seds}, the strong emission line in the SED of QSOII around $6570\mathring{A}$ in the $r$ band cannot be covered. To account for most of the features in the SED, the linear interpolation on the pixel level between each pair of adjacent simulated PSFs is applied to the PSFs.  
These mocked PSFs are produced with the step of wavelength $10\mathring{A}$. With the PSF at each wavelength and the SED templates, we generate the effective PSFs and calculate the colour bias $b_c$ (Eq.\,\ref{eq:colourbias}) between star and galaxy, and between QSO and galaxy.

In Fig.\,\ref{fig:show colour bias}, one example of the colour biases between Irr-galaxy and the three types of stars, or between Irr-galaxy and the two types of QSOs are presented. We calculate the colour bias on each grid over the whole FOV ($30\times 30$ grid) and show the magnitude distribution of $b_c$. Both the colour bias of size and ellipticity has positive and negative values. We only show their absolute value for better visibility.
The bias distributions of $R$ and $\epsilon$ show differences in all bands using star SED or QSO SED. The bias varies dramatically between bands as well, especially $b_c$ of $R$. $b_c$ of $\epsilon$\footnote{To distinguish the ellipticity of PSF or galaxy, we use ellipticity only for that of PSF, and use shear 
$\gamma$ for that of the galaxy.} is relatively large, one of the reasons is that the intrinsic ellipticity of the PSF is small $\sim 10^{-3}$. The absolute difference is $\sim 10^{-5}$, which gives the order of relative bias for ellipticity roughly $10^{-2}$. It shows less difference between the four bands. Similar distributions are shown when we use stars or QSOs to estimate PSF. Three major properties can be summarised from Fig.\,\ref{fig:show colour bias}:
(1) In general, $b_c$ of $R$ is smaller and has a narrower distribution over the FOV than that of $b_c$ of $\epsilon$. (2) the colour bias in the band of a short wavelength usually is greater than that of a longer wavelength. One can see from Fig.\,\ref{fig:filters&seds}, that there are more diversities at a shorter wavelength in the spectra of all the sources. (3) the strong emission line in SED can have a significant impact on $b_c$. For example, an emission line of QSOI around $5000\mathring{A}$ can be redshifted into $i$ band when the QSO is at $z\sim0.5$. Thus, from the panel of QSOI (0.5) one can see that the bias in the $i$ band is drastically large than that in the other cases. A similar situation is that there is an emission line of QSOII at $6570\mathring{A}$. It will be redshifted into the $z$ band when the QSO is at $z=0.5$. In additional tests, we check the difference of SED between star/QSO and galaxy. A similar trend can be found as well.

\subsection{Colour bias with colour and redshift}

\begin{figure}
\centerline{
   \includegraphics[width=0.9\textwidth]{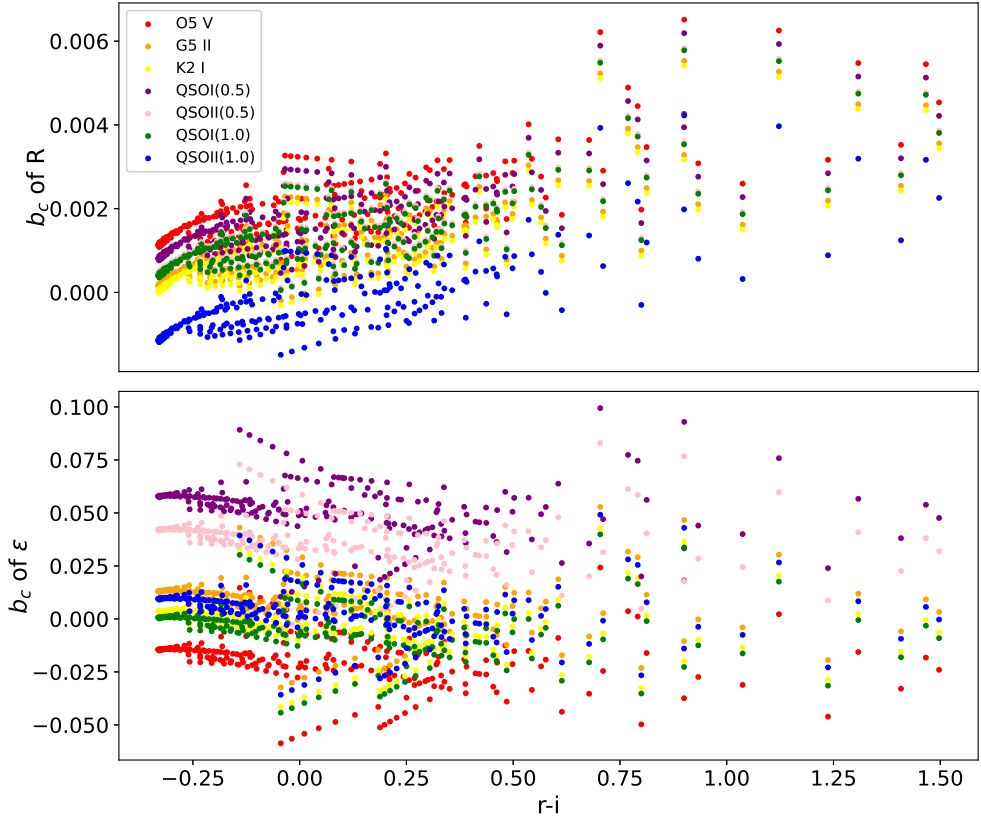}}
   \caption{Colour bias of the PSF between synthetic galaxies and stars/QSOs in the $i$ band as a function of the colour of galaxies. The red, orange and yellow points represent the bias using the star of type O5 V, G5 II and K2 I; the purple, pink, green, and blue points represent the bias using Type-I QSO (z=0.5), Type-II QSO (z=0.5), Type-I QSO (z=1.0), Type-II QSO (z=1.0) respectively. 11 synthetic galaxy SEDs between redshift $[0,2]$ with step 0.1 are used. The zero point of the colour is arbitrarily set.}
   \label{fig:psf size-colour}
\end{figure}

Since the colour bias simultaneously correlates with the SED of the galaxy and that of the star, and the colour of the source can broadly reflect the trend of the SED, the colour of the galaxy and the star can provide a rough estimate of the bias. Such a relation has been studied using real data in the Subaru Strategic Program survey \citep{2018MNRAS.479.1491C}. To study the relation for CSST, we generate a synthetic galaxy SED library. Each synthetic galaxy SED in our mock library is generated by combining an Ell SED and an Irr SED. They are used to mimic the spectrum of the bulge and disk of the galaxy. Different weighting are employed for the two components, i.e. SED$_{\rm gal}$ $=w$ SED$_{\rm Ell}$ $+(1-w)$SED$_{\rm Irr}$, where $w$ standing for the weighting. Eleven steps of $w$ from 0 to 1 are selected in our mock library, and 21 redshift steps between 0 and 2 are used. The colour $r-i$ is calculated from the integrated flux of the two neighbouring bands. Since we do not employ any particular magnitude system, the zero point of the colour index is arbitrary. The resulting $b_c$ are averaged over the FOV for each weighting $w$ and redshift combination and are shown in Fig.\,\ref{fig:psf size-colour}. Three types of stars and two types of QSOs at different redshifts ($z=0.5,1.0$) are used for the SED of PSF. The colour bias in the $i$ band is shown. A clear uptrend of $b_c$ of $R$ can be seen, i.e. the redder the galaxy, the greater the size bias. Especially for the case of a red galaxy and a blue star, the bias of $R$ can be up to $\sim 0.006$. The $b_c$ of $\epsilon$ do not have an obvious trend.

\begin{figure}
\centerline{
   \includegraphics[width=1\textwidth]{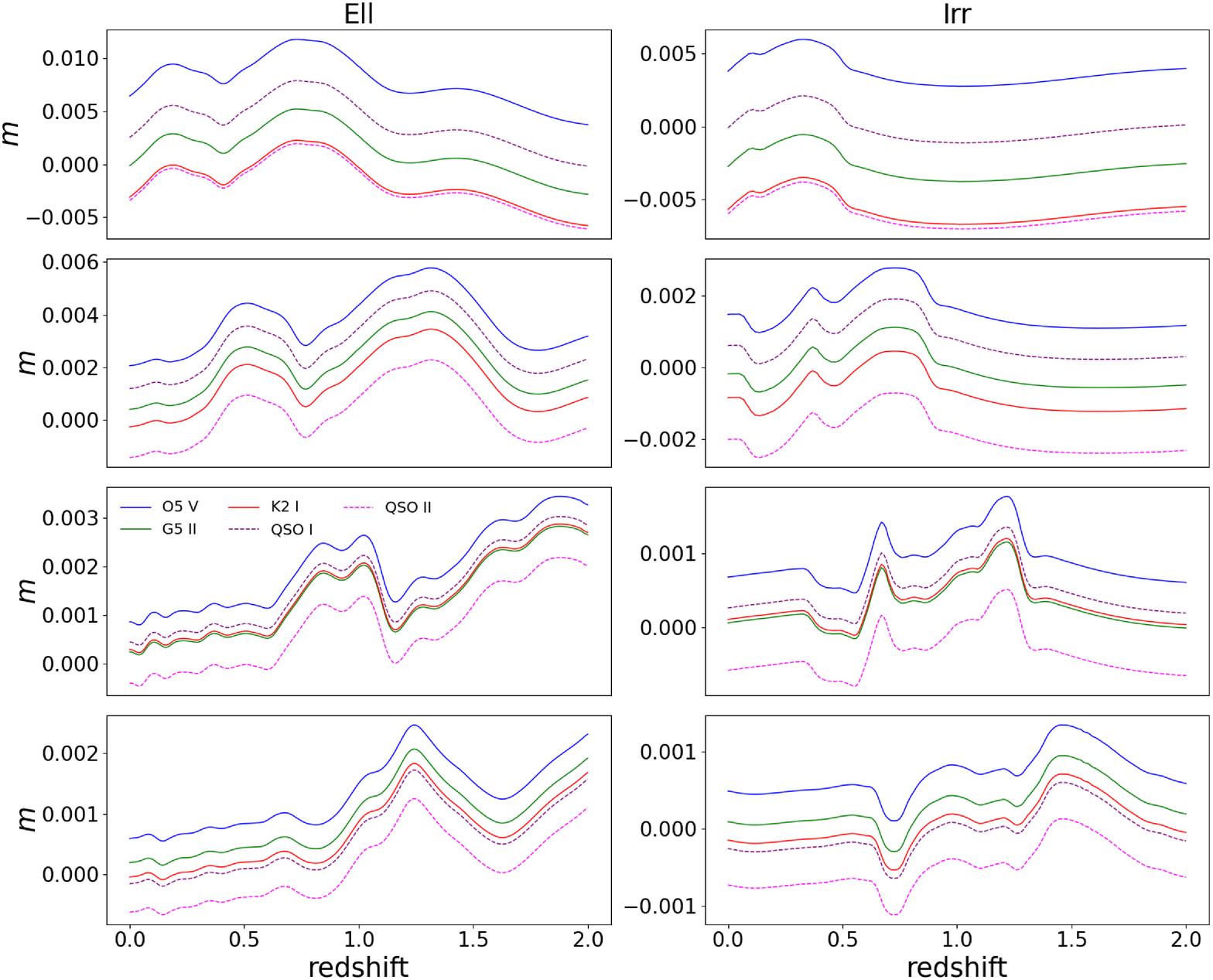}}
   \caption{The multiplicative systematic bias of shear measurement due to the colour bias vs. redshift. From top to bottom is the result in $g, r, i, z$ band respectively. In the left (right) column, the SED of an Ell (Irr) galaxy is used. Different SEDs of stars/OSQs are presented by different colour curves.}
   \label{fig:sys err vs. redshift}
\end{figure}

Moreover, the SED of the galaxies are redshifted when they are at a cosmological distance. The colour bias, therefore, changes correspondingly, especially when the SED of the galaxy has strong emission lines. The bias in PSF can propagates into the weak lensing shear measurement \citep{2010MNRAS.405..494C,2008A&A...484...67P,Massey_2012}, which is eventually important to us. We adopt the approach in \cite{Massey_2012} to estimate the multiplicative bias $m$ due to the imperfect knowledge of PSF

\be
m=2\langle\frac{R^2_{PSF}}{R^2_{gal}}\rangle\frac{\langle\delta(R^2_{PSF})\rangle}{\langle R^2_{PSF}\rangle}+\langle\frac{R^4_{PSF}}{R^4_{gal}}\rangle\frac{\sigma^2[R^2_{PSF}]}{R^4_{PSF}},
\label{eq8}
\ee
where $R_{gal}$ is the size of galaxy, $R_{PSF}$ is the size of PSF, $\delta(R^2_{PSF})$ represents the bias of $R^2_{PSF}$ from its true value and $\sigma^2[R^2_{PSF}]$ is the variance of $R^2_{PSF}$.
For simplicity, we keep $R_{PSF}/R_{gal}$ as $1/1.5$. Same as that in the previous section, we calculate the average over the whole FOV for each redshift. Three types of stars and two types of QSOs at $z=0.5$ are used as references.
We present the shear measurement error due to the colour bias in Fig.\,\ref{fig:sys err vs. redshift}.
The bias using the O5 V star is greater than the other cases in general. It agrees with our expectation since we find a strong difference in SEDs between O5 V and galaxies. 
Two main features can be seen here: 1) The order of magnitude of the shear bias is similar between the two types of galaxies. 2) the amplitude of shear bias decreases with the band, i.e. $b_c^g>b_c^r>b_c^i>b_c^z$. We can see that the multiplicative systematic bias in shear measurement caused by colour bias is smaller than typically current constraint $m\sim10^{-2}$  \citep{2016MNRAS.460.2245J,2018MNRAS.481.3170M,2023A&A...670A.100L}. 
However, Eq.\,\ref{eq8} only includes the contribution of second-order moments. As pointed out by \cite{2022MNRAS.510.1978Z,2023MNRAS.520.2328Z}, the higher order moments of the PSF can be a significant source of the shear bias in the upcoming surveys. The exact magnitude of such an effect requires sophisticated simulations, which will be left in future work.

\begin{figure}
\centerline{
   \includegraphics[width=1.0\textwidth]{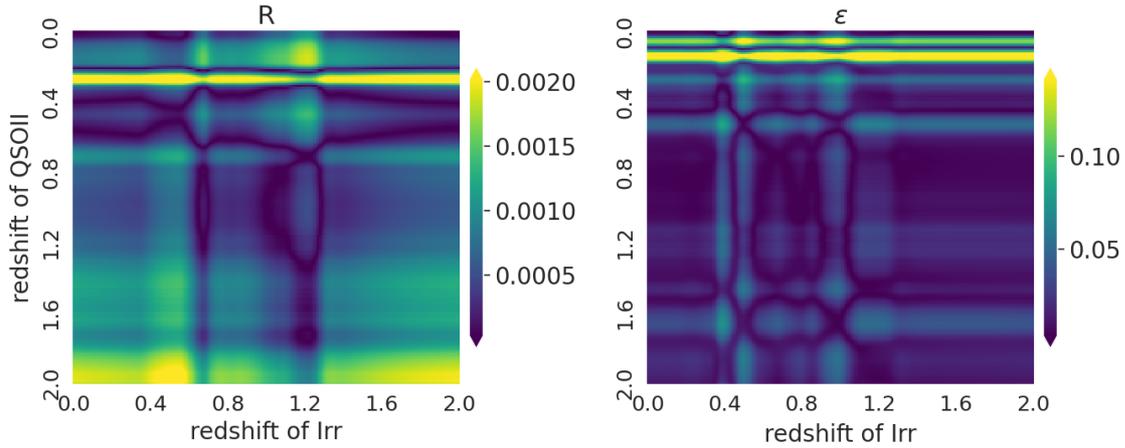}}
   \caption{The heat map of the colour bias between Irr and QSOII in the $i$ band. The horizontal (vertical) axis is the redshift of the galaxy (QSO). From left to right are the bias of size and the bias of ellipticity. The redshift of the galaxy or QSO is range from $0$ to $2$ with step $0.01$.}
   \label{fig:Irr-QSOII vs. z}
\end{figure}

The stars we used to estimate PSF are local, i.e. $z=0$, while the QSOs have a wide distribution of redshift. To see the colour bias with different combinations of QSO redshift and galaxy redshift, we perform an additional test with the redshift range from $0$ to $2$ with a step of $0.01$ for both galaxies and QSOs. Fig.\,\ref{fig:Irr-QSOII vs. z} shows the colour bias between Irr and QSOII. 
From Fig.\,\ref{fig:Irr-QSOII vs. z}, one can see that $b_c$ of $R$ is rough $10^{-3}$ and $b_c$ of $\epsilon$ is about $0.03$ in the most redshift combinations (relatively dark region). The colour bias does not change drastically in different redshift combinations of the galaxy and the QSO in general. However, there are some vertical and horizontal bright stripes appearing on both left and right panels, which shows acute changes of the colour bias at these redshift combinations, especially when the QSO at $z\sim0.27$ (left panel, bias of $R$) and $z\sim0.05$ or $0.14$ (right panel, bias of $\epsilon$). It is due to the strong emission line of QSOII (at around $6570\mathring{A}$) in $i$ band at these redshifts. It can give a large weighting of the PSF at this wavelength. The wavelength dependence of PSF size or PSF ellipticity is different, thus the stripe appears at different redshifts on the left or right panel.

\begin{figure}
   \centerline{
   \includegraphics[width=0.9\textwidth]{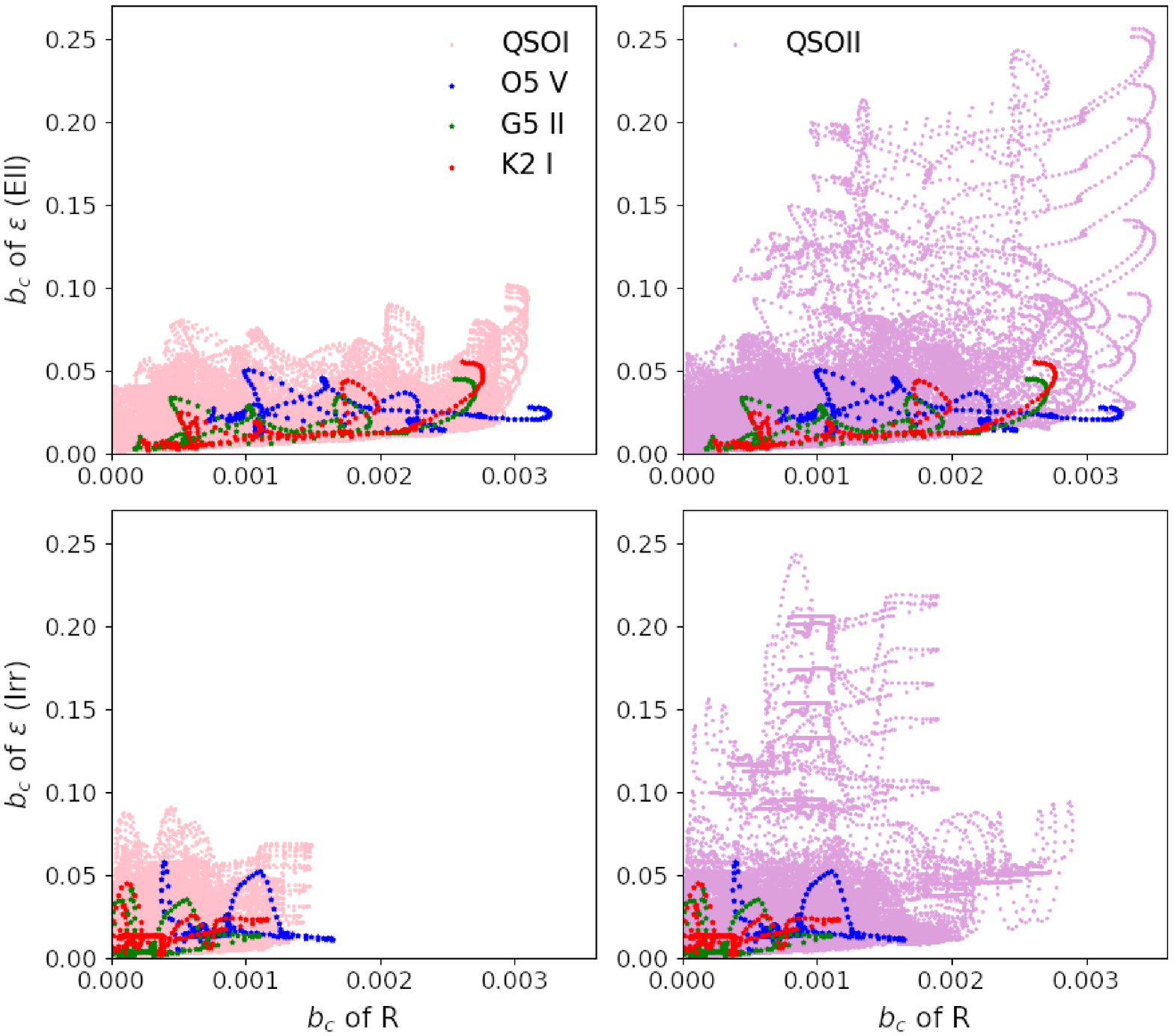}}
   \caption{
   The distribution of colour bias in $i$ band. Each point represents the colour bias of $R$ (horizontal axis) and bias of $\epsilon$ (vertical axis) for one case of combination, star with galaxy or QSO with galaxy. Different redshifts for galaxy and QSO are used. In the left (right) panels, we compare that of the stars with QSOI (QSOII). In the top (bottom) panels, the colour bias is calculated with respect to the Ell (Irr) galaxy. Three different colours present the bias using three SED of stars.}
   \label{fig:colour bias vs. z (scatter)}
\end{figure}

We further compare the colour bias between using star or QSO as reference. The same redshift range is adopted for QSO and galaxy. The result is shown in Fig.\,\ref{fig:colour bias vs. z (scatter)}. 

We find that the bias of ellipticity between the galaxy and QSO is bigger than that between the galaxy and star, especially for QSOII. The colour bias of size with stars is similar to that with QSO. For both two kinds of galaxies (Ell or Irr), the colour bias using QSOII is larger than using QSOI. The main reason for that is the strong emission line and the shape of the spectrum of QSOII. The difference between biases using QSO or star can be explained by the same reason.

\subsection{Calibration of colour bias}

\begin{figure}
\includegraphics[width=0.85\textwidth]{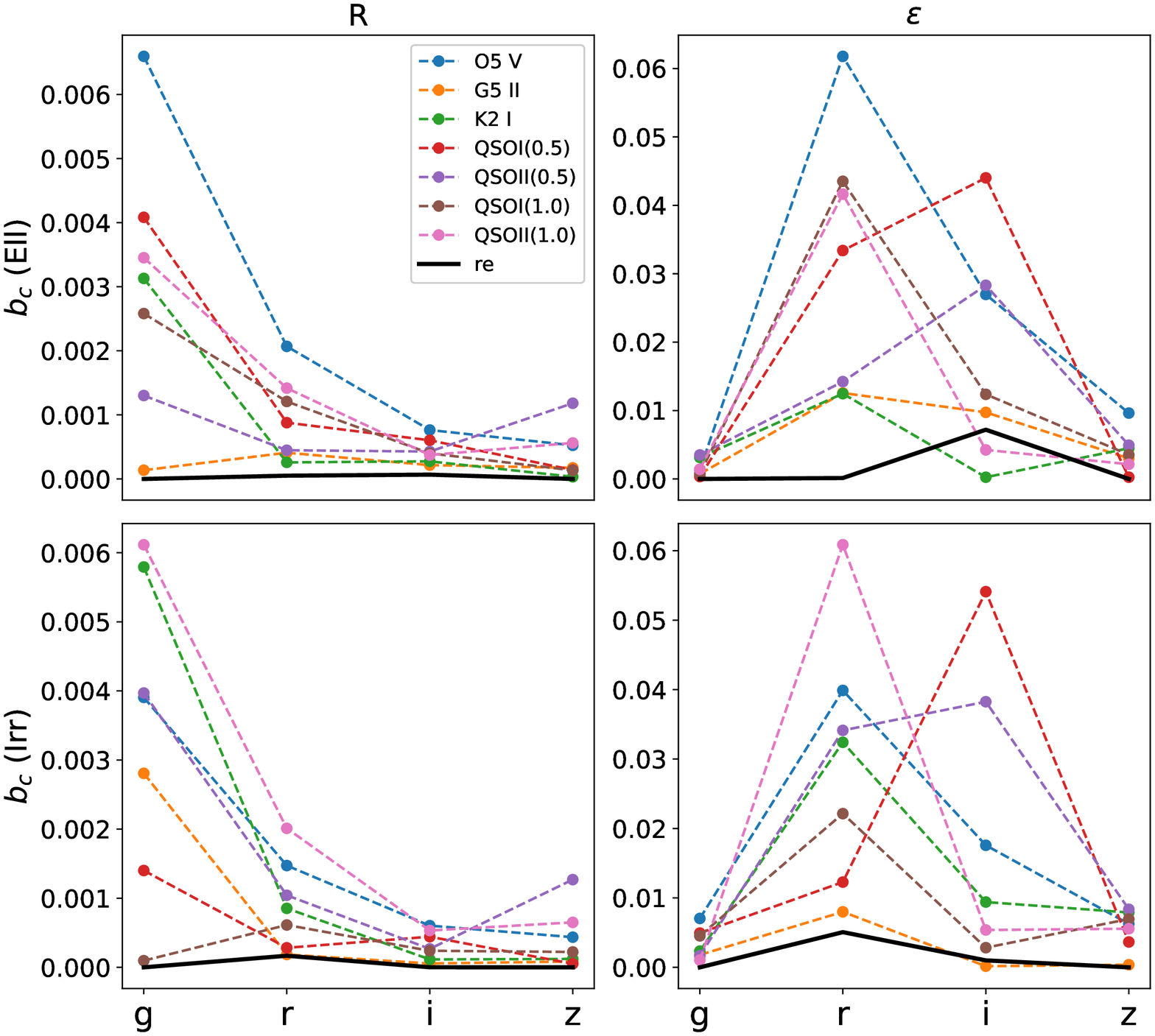}
   \caption{The mean colour bias over the FOV in four bands of CSST. From left to right is $b_c$ of $R$ and $b_c$ of $\epsilon$. The Irr is used as the reference. The black solid lines present the bias using reconstructed SED of the galaxy.}
\label{fig:linear_interpolation}
\end{figure}

The colour bias can be calibrated from the SED or the colour of the galaxy and star\citep[e.g.][]{2018MNRAS.477.3433E}. In this part, we present the calibration of the colour bias. In order to isolate the colour bias, the other noises, such as sky background or readout noise are not included here. There are possible couplings between the colour bias and the other noises. But the colour bias has systematic dependence on the SED of the star and galaxy. We use linearly interpolating from the two neighbouring bands of the galaxy photometry to reconstruct the SED of galaxy \cite{2013MNRAS.432.2385S}. The morphology of the galaxy are not included. We follow the same procedure that is used in Section 3.1 and obtain the reconstructed PSF (PSF$^{re}$ for short) from the interpolated SED, then compare it with PSF$^{gal}$. For $r$ bands, one can interpolate the SED either from $g, r$ bands or from $r, i$ bands. We reconstruct the SED from both ways to calculate the colour bias, and show the average colour bias from two reconstructions. The same procedure is applied for the $i$ band.
In Fig.\,\ref{fig:linear_interpolation}, the colour bias using reconstructed SED (black solid lines) is compared with the other cases. As one expects, there is a significant improvement, i.e. the biases using the reconstructed SED are smaller than the others in most cases. 
One weakness of this method is that it cannot recover the acute structure in the galaxy SED, such as emission lines in the Irr galaxy. 
Moreover, the PSF model of wavelength dependence is required in this method.

\section{Colour gradient bias}
\label{sect:CGbias}

The SED of a galaxy varies spatially, generating colour gradient. We estimate the shear measurement bias when the colour gradient of the galaxy is ignored, i.e. the CG bias. It has been shown that the CG bias depends on several factors, such as the measurement method, the relative size of the galaxy with respect to the size of the PSF, the pixel scale of the CCD, the transmission function of the filters etc \citep[e.g.][]{2012MNRAS.421.1385V,2013MNRAS.432.2385S}. On top of that is the colour gradient of the galaxy. For example, if there is a great difference in the SED between the bulge and disk of the galaxy, one would expect a significant CG bias. More detailed discussions can be found in \citet{2013MNRAS.432.2385S}. We evaluate the CG bias for CSST four filters following a similar approach in \cite{2018MNRAS.476.5645E}, and only introduce basic steps here (one can find more details from Fig.\,1 in \cite{2018MNRAS.476.5645E}). Following the step in the Figure, one can generate the galaxy images without CG bias from the top flowchart: 1) convolve the image with PSF at each wavelength; 2) integrate the images over the band of wavelength; 3) deconvolve the image by the effective PSF; 4) shear the deconvolved image.
In the bottom flowchart, one follows the real image process and can simulate the image with CG. The only difference is that the shear step is at the beginning. In reality, there are convolution with the effective PSF in the end for both top and bottom flow. Since we don't employ any particular PSF correction method, a direct deconvolution is used, and cancel out the last convolution. Then we use the second-order brightness moment and Gaussian weighting function to measure the shear from these two kinds of images.

We combine a bulge and a disk to mimic the spatial variation of synthetic galaxy SED. Again we use an Ell galaxy SED for the bulge and an Irr galaxy SED for the disk. The image profiles are described by the Sersic profile at each wavelength
\be
I_s(x,y) = I_0 {\rm e }^{-\kappa(\frac{\theta}{r_h})^{1/n}},\qquad{\rm with \qquad}
\theta=\sqrt{x^2/q+y^2q},
\ee
where $I_0$ is the central intensity, $\kappa=1.9992n-0.3271$, $n$ is the Sersic index, and $q$ is the axis ratio, $r_h$ is the half-light radius. We take $n=1.5$ for the bulge and $n=1$ for the disk. The source galaxies are assumed to be circular initially. Two sizes of galaxies are considered in the first test. The size for the big galaxy and the small galaxy is given by the half-light radius of its bulge and disk: $r_h^{bulge}=0.17, r_h^{disk}=1.2$  arcsec and $r_h^{bulge}=0.09, r_h^{disk}=0.6$  arcsec respectively. The total flux of synthetic galaxy is normalised at the wavelength of $550$ nm where the bulge and disk contain 25 and 75 percent of the flux respectively. A shear value ($\gamma_1=\gamma_2=0.05$) is uniformly used in all the tests. The PYTHON-based GalSim package \citep{2015A&C....10..121R} is used to simulate the galaxy image, which has been widely adopted in weak lensing studies \citep[e.g.][]{Hoekstra16}. The stamp size of each galaxy image is $512\times512$ pixels with pixel size $0.074$ arcsec. The wavelength sampling interval is $1$ nm. The PSF model is simply simulated by the Airy function, which shows a similar shape to the PSF in the CSST $i$ band \citep{2022AJ....164..214S}. The profile of Airy disk is expressed as
\be
P(x)=\frac{I_0}{(1-\kappa^2)^2}
\rund{\frac{2J_{1}(x)}{x}-\frac{2\kappa{J_{1}({\kappa}x)}}{\kappa{x}^2}}^2,
\label{eq:define_AriyPSF}
\ee
where $I_0$ is the maximum intensity at the centre, $\kappa$ is the aperture obscuration ratio, and $J_1(x)$ is the first kind of Bessel function of order one; $x$ is defined as $x=\frac{\pi\theta}{\lambda{D}}$. We take the CSST aperture diameter $D=2$ m and obscuration $\kappa=0.1$ in our simulation. In this study, we do not consider the evolution of the galaxies over the redshift, i.e. we use the same angular half-light radius of the bulge and the disk for galaxies at different redshifts. But the size of synthetic galaxies vary due to the SEDs, i.e. the ratio of the bulge-to-disk can change with redshift. The spatial variation of CG bias over the FOV has not been taken into account.

\begin{figure}
\centerline{\includegraphics[width=0.85\textwidth]{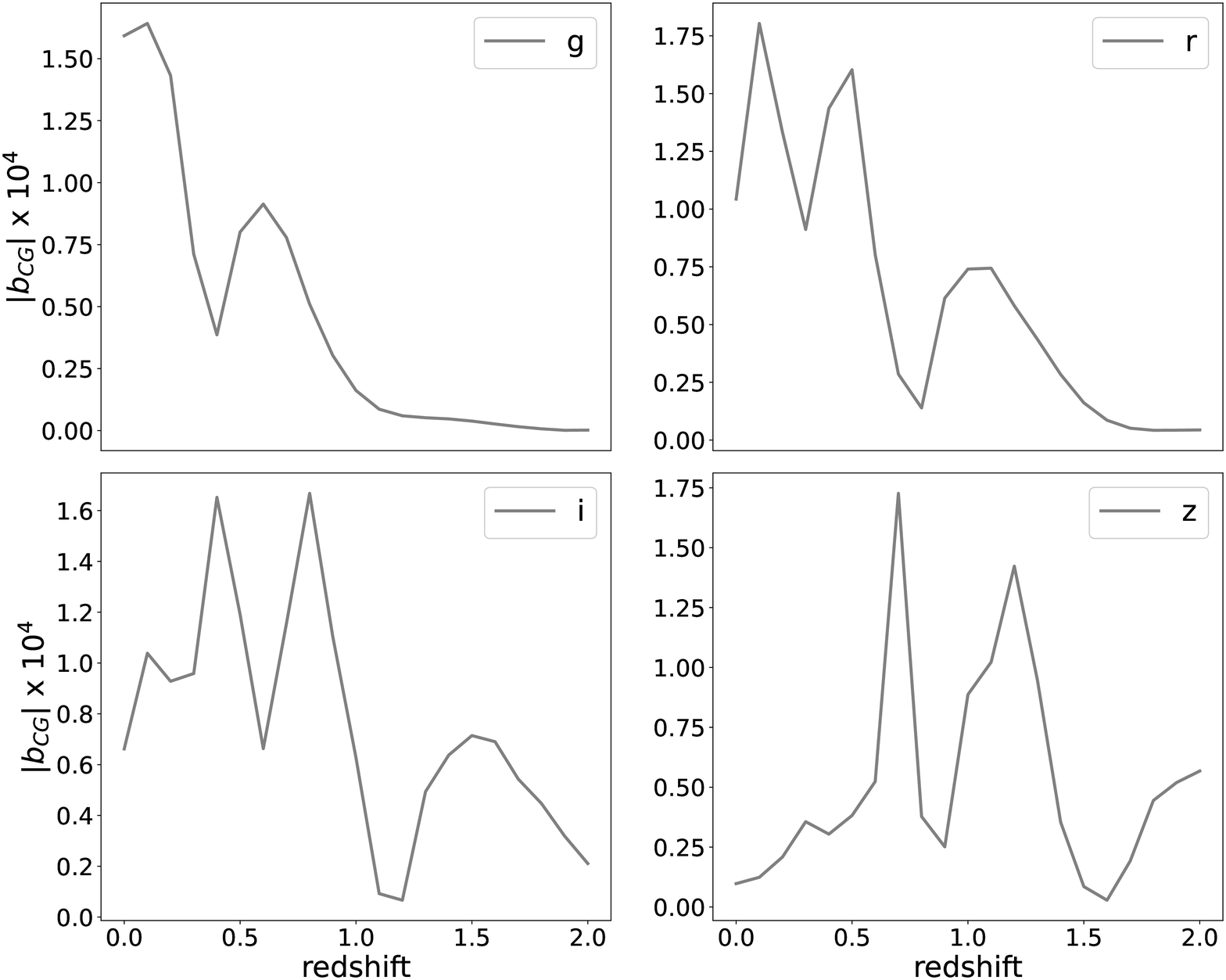}}
\caption{The colour gradient bias as a function of redshift for the 'big' galaxy in $g, r, i, z$ band of CSST.}
\label{fig:CG_bias_vs_redshift(B)}
\end{figure}

\begin{figure}
\centerline{	\includegraphics[width=0.85\textwidth]{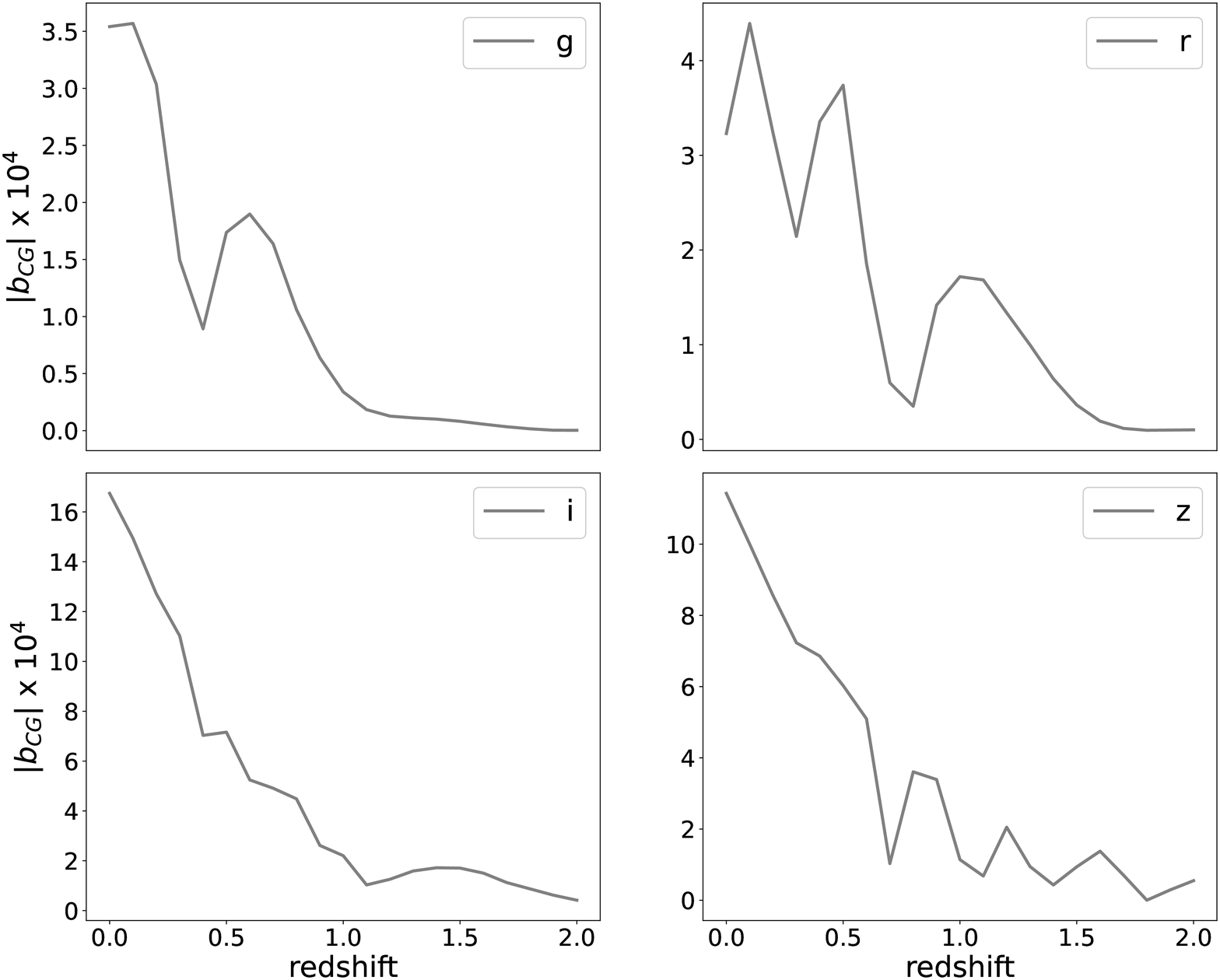}}
\caption{Same as Fig.\,\ref{fig:CG_bias_vs_redshift(B)} but for the 'small' galaxy.}
\label{fig:CG_bias_vs_redshift(S)}
\end{figure}

We calculate the ``shear'' from the simulated images, and obtain the bias $b_{CG}$. In Fig.\,\ref{fig:CG_bias_vs_redshift(B)} and Fig.\,\ref{fig:CG_bias_vs_redshift(S)}, we show the $b_{CG}$ vs. redshift in each band with step $\Delta z=0.1$. Overall, the bias is small, especially in the big galaxy (\,$<1.8\times10^{-4}$\,). It is slightly larger in the small galaxy (\,$<1.7\times10^{-3}$\,). For both galaxies, the bias decreases with redshift, which agrees with the previous study \cite{2018MNRAS.476.5645E}.
The two peaks in the CG bias curves correspond to the two strong emission lines of Irr SED. 
We estimate the colour gradient of our synthetic galaxy for comparison. In order to calculate the 'colour', we split each band into two at the central wavelength of the bandwidth. The flux ratio between the two sub-band of the galaxy image is taken as the colour. The colour difference between the central region
($r < r^{bulge}_h$) and outer region ($r^{bulge}_h < r < r^{disk}_h$) of the galaxy image is defined as the colour gradient. We calculate the colour gradient as a function of the redshift and find this relation shows a similar trend as that of CG bias.


\begin{figure}
\centerline{	\includegraphics[width=1\textwidth]{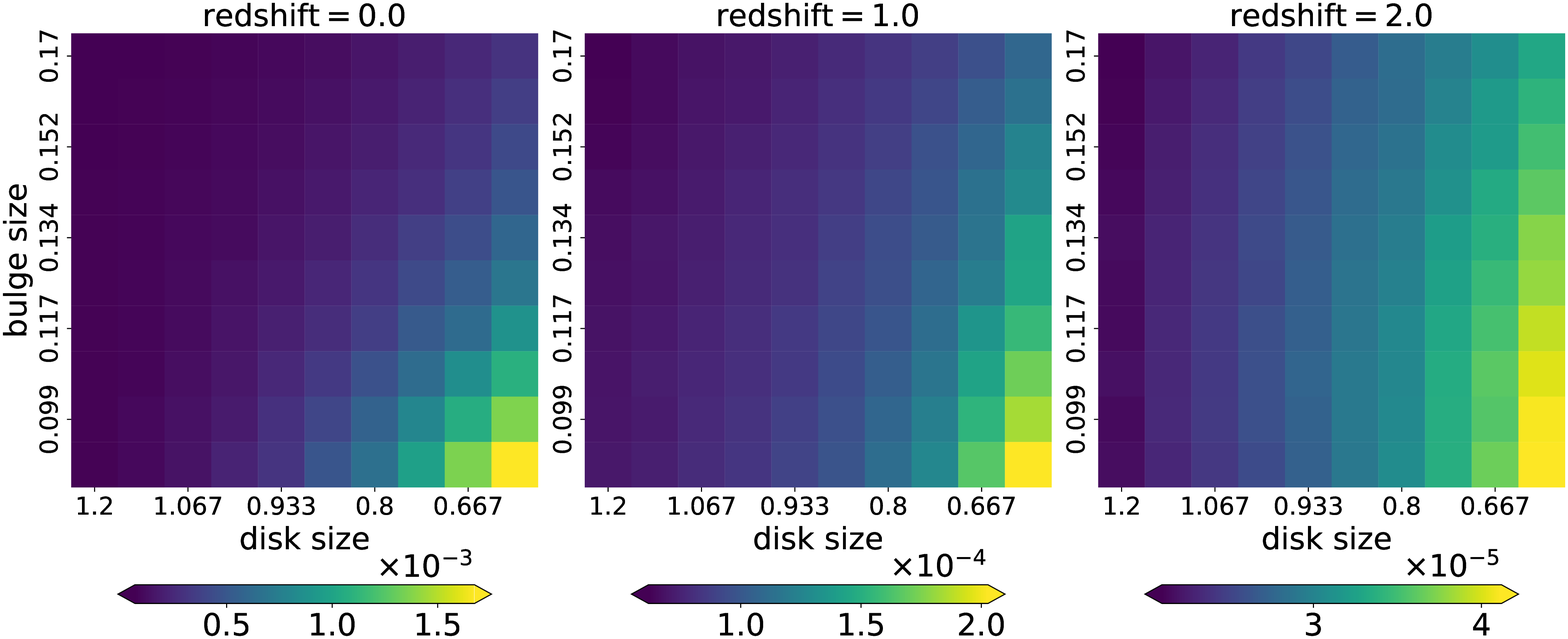}}
\caption{The relation between the colour gradient bias and the galaxy size in the CSST $i$ band. From left to right, three SEDs corresponding to different redshifts have been used. The different size combinations of the bulge and disk are used in each panel. The unit of the size is arcsec and is presented by the half-light radius. The size of the bugle decreases from $0.17$ arcsec to $0.09$ arcsec, and the disk size decreases from $1.2$ arcsec to $0.6$ arcsec.}
\label{fig:CG_bias_vs_gal_size}
\end{figure}

In the $i$ band and $z$ band, there are unusually large biases in the case of a small galaxy at low redshift. The main reason is that the relative size of the galaxy with respect to that of PSF becomes small and critical.
From the relative strength of SED between the Ell and Irr galaxies (Fig.\,\ref{fig:filters&seds}), one can see that the synthetic galaxy is bulge dominated and compacted at low redshift, and becomes disk dominated and extended at high redshift.
In additional tests with smaller PSF sizes, the CG bias shows similar behaviour as that using large galaxies. 

Moreover, we take different combinations between bulge size and disk size into account. Only CSST $i$ band is performed in this part. Fig.\,\ref{fig:CG_bias_vs_gal_size} shows the relation of CG bias with the size of two components of the synthetic galaxy. We use the bulge size from $0.17$ arcsec to $0.09$ arcsec and the disk size from $1.2$ arcsec to $0.6$ arcsec in each panel. Three redshifts ($z=0.0, 1.0, 2.0$) are used here. We can see that the CG bias is more sensitive to the bulge size than the disk size. There is rapid growth when the galaxy becomes small.

\section{Discussion and Summary}
\label{sect:summary}
Weak gravitational lensing is one of the most powerful tools in modern cosmology. The next generation weak lensing surveys, such as $Euclid$, CSST, LSST and WFIRST, will measure the weak lensing signal with unprecedented precision. The shape measurements benefit from the compact diffraction-limited PSF of the space-borne telescope. Thus small instrumental effects can become important and need to be carefully investigated. One of the effects is the wavelength dependence of the PSF. In this paper, we study two biases due to the chromatic PSF in the CSST weak lensing measurement. The first one is the colour bias. The difference between the star SED and galaxy SED leads to the difference between PSF obtained from star images and the PSF that smears the galaxy. The other one is the colour gradient bias. It arises due to the spatial variations in the colours of galaxies. In our simulations, we find both of these two biases are small in most cases of the weak lensing measurement in the four optical filters of CSST. 

For the colour bias, two types of SEDs are taken as reference for the galaxy. One is an irregular galaxy, and the other one is an elliptical galaxy. Three types of SEDs of the star (O5 V, G5 II, K2 I) and two types of SEDs of QSO (type I and type II) are used since QSO has point like image and can be misclassified as the star. We study the colour bias of PSF size and ellipticity for the CSST. 
We find that the colour bias of size is one order of magnitude smaller than that of ellipticity in general. The order of magnitude distributions of the colour bias of ellipticity show similarity in four bands, and cover a wide range. One of the reasons is that the ellipticity of the PSF is sensitive to the position on the field of view. While the distribution of the colour bias of size is centrally distributed in all bands and has widely separated among the filters. Two other properties can be found: 1) in the shorter wavelength filter, the colour bias will be larger; 2) the strong emission line of the source galaxy has a big impact on the colour bias.

We study the dependence of colour bias on some other factors. By generating a synthetic galaxy SED library, we find that there is a positive correlation between the colour bias of size and the colour of the galaxy. While the relation between the bias of ellipticity and galaxy colour is not clear.
In the test of colour bias for the galaxy at different redshifts, we do not find a large difference between the elliptical galaxy and the irregular galaxy. We also calculate the shear measurement error due to colour bias, and find in all four bands, the shear measurement errors are small ($<1\%$).
In the redshift combination of galaxy and QSO, we find the colour bias is significantly affected by the strong emission lines in the SED of galaxy and QSO. 
We perform calibration to the colour bias by reconstructing the SED of the galaxy. We use the brightness of two neighbouring bands and linearly interpolate the SED of the galaxy. The PSF based on the reconstructed SED provides a better estimate of PSF which is close to one that smears the galaxy.

For the colour gradient bias, only the multiplicative bias is studied. We adopt the Airy profile to simulate the PSF and estimate the CG bias at different redshifts for four CSST bands. For galaxies at different redshifts, we only consider the changing of the SED, but the evolution of the galaxies. The CG bias is also small $\sim 10^{-4}$ and decreases with redshift in general. Similar to that in the colour bias, a strong emission line of the source galaxy can increase the CG bias.
When the size of the galaxy becomes small, the CG bias also increases rapidly. Particular in the CSST $i$ band and $z$ band, the bias can be up to $1.7\times10^{-3}$ and $1.2\times10^{-3}$ respectively. The main reason is that the ratio between the PSF size and galaxy size is relatively large in these bands. Our results show that the CG bias in CSST shear measurement is a subdominant effect for relatively big galaxies, but for small galaxies, in CSST $i$ and $z$ band CG bias can cause non-negligible systematic bias in the Stage-IV weak lensing survey.

In this study, we show that the chromatic effect in PSF can be small in most cases of shear measurements. However, some exceptions one has to be careful of. For example, the misclassification of QSO as the star. Measuring shear from small-size galaxies. Moreover, there are some simplifications in our study. First, in our simulation, we only consider the optics of the CSST. Other instrumental effects, such as those from the detector need to be included as well. 
The linear reconstruction method to calibrate the colour bias needs to be processed for every galaxy, which will cost massive computations. The relation between the colour and the colour bias can provide a neat estimate of the bias.
The evaluation of the colour gradient bias in this paper is based on the Airy profile. An estimate with realistic noise and calibration to other wider band filters \citep{2023A&A...669A.128L} will be important to the weak lensing measurement as well.

\begin{acknowledgements}
We would like to thank the referee for the comments and suggestions on our draft.
This work was funded by the National Natural Science Foundation of China (NSFC) under No.11873006, 11933002, 11903082, U1931210. We acknowledge the science research grants from the China Manned Space Project with No.CMS-CSST-2021-A01, No.CMS-CSST-2021-A12, No.CMS-CSST-2021-A07.
\end{acknowledgements}


\bibliographystyle{raa}
\bibliography{lens,cgbibtex}

\label{lastpage}
\end{document}